\documentstyle[preprint,tighten,aps,12pt,epsf]{revtex}

\begin{document}

\hfill INT \# DOE/ER/40561-294-INT96-00-151

\begin{center}
{\large \bf
Spin and Weak Interactions in Atoms and Nuclei\footnote{Plenary Talk,
12th International Symposium on High Energy Spin Physics (SPIN 96),
Amsterdam, 10-14 September, 1996}
\\ }
\vspace{5mm}
M.J. Musolf\footnote{National Science Foundation Young Investigator}
\\
\vspace{5mm}
{\small\it
Institute for Nuclear Theory, Box 351550 \\
University of Washington, Seattle, WA 98195 U.S.A.
\\ }
\end{center}

\begin{center}
ABSTRACT

\vspace{5mm}
\begin{minipage}{130 mm}
\small
The use of spin observables to study the semi-leptonic and non-leptonic
weak interaction in atoms and nuclei is surveyed. In particular, the use
of semi-leptonic neutral current scattering and atomic parity violation
to search for physics beyond the Standard Model is reviewed. The status
of nuclear parity violation as a probe of the weak N-N interaction is
surveyed. Possible atomic and nuclear signatures of parity conserving, 
time-reversal violating interactions are also discussed.
\end{minipage}
\end{center}

The use of atomic and nuclear processes to eludicate the structure of
the weak interaction has a long and illustrious history. With the
advent of very high-precision, high-energy studies at LEP, SLC, and
the Tevatron, it is natural to ask what role, if any, low-energy weak
interaction studies might continue to play in uncovering new aspects
of electroweak physics. In this talk, I wish to focus on three areas
in which such a role can be envisioned: (a) constraining possible
extensions of the Standard Model in the neutral current (NC) sector;
(b) probing the strangeness-conserving non-leptonic weak interaction;
(c) searching for signatures of interactions which conserve parity
invariance but violate time-reversal invariance. In each case, I
will emphasize the insight which might be derived from the analysis
of spin-observables.

\medskip
\noindent{\bf Neutral Current Studies}
\medskip

Although $Z$-pole observables from LEP and the SLC are placing
increasingly tight constraints on possible extensions of the
Standard Model (SM), there still exists a window of opportunity for
low-energy observables. To illustrate, one may consider three
different types of \lq\lq new physics" which may appear in NC
interactions: (a) additional neutral
gauge bosons, (b) effective interactions arising from lepton
and quark compositeness, and (c) additional heavy
physics which modifies the SM vector boson propagators. While
extensions of types (a) and (b) -- known as \lq\lq direct" --
contribute at tree level, those of type (c) arise via loops and
are correspondingly referred to as \lq\lq oblique".
Insofar as new direct interactions are associated
with mass scales differing from $M_Z$, the high-energy $e^+e^-$
accelerators will be rather transparent to their presence.
In contrast, $Z$-pole studies place non-trivial constraints
on oblique corrections, since the latter modify the corresponding
observables.

The presence of additional, neutral gauge bosons is expected within
the context of a variety of grand unified theories in which some
group ${\cal G}$ associated with an un-broken gauge symmetry at
a high mass scale spontaneously breaks down to the SU(3$)_c\times$SU(2$)_L
\times$U(1$)_Y$ symmetry of the weak scale. A particularly useful scheme
for considering the generation of additional neutral gauge bosons is
associated with the group $E_6$, which arises naturally within the
context of heterotic string theory [1]. The various scenarios for $E_6$
breakdown can be characterized by assuming that at least one relatively
low-mass neutral boson $Z'$ generated in the process. Moreover, one
may decompose this $Z'$ as
\begin{equation}
Z'=\cos\phi Z_\psi+\sin\phi Z_\chi
\end{equation}
where $\psi$ and $\chi$ denote the U(1) groups appearing in the
breakdowns $E_6\to SO(10)\times U(1)_\psi$ and $SO(10)\to
SU(5)\times U(1)_\chi$. Different scenarios for the breakdown of
$E_6$ are reflected in different values of the angle $\phi$. 

From the standpoint of low-energy NC observables, the $Z_\chi$ is
the most interesting. The reason is that the $Z_\psi$ has only
axial vector couplings to the known leptons and quarks. In the
limit that the $Z-Z'$ mixing angle vanishes, it therefore
cannot contribute to atomic PV or PV electron scattering, which 
are my focus here. I consider only the case where the
$Z'$ does not mix with the SM $Z$, since results from $Z$-pole measurements
place stringent constraints on $M_{Z'}$ for mixing angle $\theta_{ZZ'}$
differing non-negligibly from zero. For $\theta_{ZZ'}\approx 0$, the
$Z$-pole observables are relatively insensitive to the presence of
a low-mass $Z'$. In contrast, low-energy observables 
offer the possibility of constraining the mass of an un-mixed $Z'$.
Direct searches for such a $Z'$ have been reported
by the CDF collaboration, yielding a lower bound on $M_{Z'}$ of
between 500 and 585 GeV for the $Z_\chi$ [2]. High-precision low-energy
semi-leptonic PV measurements could increase this lower bound by 
nearly a factor of two.

Just as $Z$-pole observables are transparent to the presence of an
un-mixed $Z'$, they are also fairly insensitive to the existence of
effective four fermion interactions arising from the assumption that
leptons and quarks are composite particles. These effective interactions
have the form [3]
\begin{equation}
{\cal L}_{\hbox{composite}} = {4\pi\over \Lambda^2} \bar{f}_1 \Gamma f_2
\bar{f}_3 \Gamma f_4,
\end{equation}
where the $f_i$ are fermion fields and $\Lambda$ is a mass scale
associated with compositeness. Current bounds from atomic PV suggest
that the corresponding distance scale $\Lambda^{-1}$
is less than 0.01 of the compton
wavelength of the $Z$. Future improvements in the precision of atomic
PV or PV electron scattering could improve this bound by more than 50\%.
These expectations compare favorably with prospective limits attainable
at the Tevatron [4].

The oblique corrections arising from modifications of the $Z$ and $W$
propagators are conveniently characterized in terms of two parameters,
$S$ and $T$ [5]. Physically, the former is associated with the presence of
degenerate heavy physics, such as an additional generation of fermions
in which the members of an iso-doublet have the same mass. The 
parameter $T$ signals the presence of weak isospin-breaking heavy 
physics, such as a non-degenerate pair of new heavy fermions. 
Most significantly, the present global constraints on $S$ favor a
central value which is slightly negative, in contrast with the predicition
of standard technicolor theories [6]. 
Both atomic PV and PV electron scattering
from a $(J^\pi, I)=(0^+,0)$ nucleus are essentially sensitive to $S$ and
manifiest only a slight sensitivity to $T$. In this respect they contrast
with most other electroweak observables, from the $Z$-pole on down. 
The present constraints from atomic PV are consistent with negative
values for $S$. Although, by themselves, these results do not significantly
affect the 68\% or 95\% CL contours of the global fits, future measurements
with improved precision could impact the location of the central values 
for $S$ and $T$.

Atomic PV and PV electron scattering are sensitive to these three examples
of SM extensions via their dependence on the so-called \lq\lq weak 
charge" of the nucleus, $Q_W$. The weak charge can be decomposed as
follows:
\begin{equation}
Q_W=Q_W(\hbox{SM})+\Delta Q_W(\hbox{new}) + \Delta Q_W(\hbox{had}) 
\ \ \ ,
\end{equation}
where \lq\lq SM" denotes the contribution arising within the framework
of the SM, \lq\lq new" represents corrections arising from SM extensions
as outlined above, and \lq\lq had" indicates corrections that depend on
hadronic and nuclear structure. In the case of atomic PV, $Q_W$
enters the PV amplitude $A_{PV}$ arising from the electron's axial
vector NC interacting with the vector NC of the nucleus. The most
precise limits on $Q_W$ have been obtained for atomic cesium, in
which $A_{PV}$ is extracted from a PV $6s\to 7s$ transition in the
presence of Stark-induced level mixing. The PV amplitude for this
transition can be written as [7]
\begin{equation}
A_{PV} = Q_W {\vec\epsilon}\cdot\langle F' M_{F'}|{\vec\sigma}|F M_F\rangle
\ \ \ ,
\end{equation}
where $F M_F$ {\em etc.} denote atomic hyperfine levels and $\vec\sigma$
is the spin of a valence electron. From the PV transition between
states having the same hyperfine quantum numbers, one extracts the
ratio
\begin{equation}
|A_{PV}/A_{\hbox{Stark}}|=\xi Q_W\ \ \ ,
\end{equation}
where $\xi$ as an atomic structure-dependent quantity. As the value of
this quantity $\xi$ requires input from atomic theory, one encounters
a theoretical, as well as experimental, uncertainty in the corresponding
value of $Q_W$. It may well be that the primary challenge for improved
constraints on $Q_W$ from atomic PV is the theoretical uncertainty.

In the case of PV electron scattering, the relevant observable is
the helicity-difference 
\lq\lq left-right" asymmetry $A_{LR}$, which may be expresssed as [8]
\begin{equation}
A_{LR} = {N_+ - N_-\over N_+ + N_-} = a_0 |Q^2|\left\{Q_W+ F(Q^2)\right\}
\ \ \ ,
\end{equation}
where $N_+$ ($N_-$) denotes the number of scattered electrons for an
incident beam of positive (negative) helicity electrons, $a_0$ is a constant
depending on the Fermi constant and EM fine structure constant, and
$F(Q^2)$ is a term dependent on nuclear form factors. Note that the
term containing $Q_W$ is nominally {\em independent} 
of hadron or nuclear structure.
In fact, there do exist structure-dependent corrections, contained in
$\Delta Q_W(\hbox{had})$, associated with
two-boson exchange dispersion corrections. 
The scale of these corrections has yet to be evaluated
reliably by theorists. The strategy for extracting $Q_W$ from $A_{LR}$
is to perform a kinematic separation of the two-terms in Eq. (6) by
exploiting the $Q^2$-dependence of the second term. One therefore 
requires sufficiently reliable knowledge of the form factors in order
to successfully carry out this program.

How well might future $Q_W$ determinations from either of these
processes do in constraining new physics? To illustrate, I will
use cesium atomic PV and PV electron scattering from a $(0^+,0)$ nucleus,
which appear to be the best cases from a variety of standpoints. The
present results for $Q_W(\hbox{Cs})$, for which the combined experimental
and theoretical error is about 2.5\%,
constrain the mass of the $Z_\chi$
to be greater than about 0.5 TeV, a limit roughly comparable with 
the bounds from the Tevatron. A future 1\% determination would push
this bound to 0.8 TeV. Similarly, a 1\% measurement of $Q_W(0^+,0)$
from PV electron scattering would yield a bound of 0.9 TeV (the
difference between the two processes follows from the different 
$u$- and $d$-quark content of the respective nuclei). In terms of
compositeness, the present cesium results require $\Lambda > 10$ TeV.
A 1\% determination of $Q_W(\hbox{Cs})$ or $Q_W(0^+,0)$ would place
this lower bound at about 16 TeV. Finally, a factor of four improvement
in the cesium precision would shift the global central value of $S$ by a factor
of four or so, assuming the same central value of $Q_W$ is obtained in
a future experiment [6]. A similar statement applies to extractions of $Q_W$
from PV electron scattering. In short, it is apparent that pushing for
improved precision in these low-energy processes could yield significant
constraints on SM extensions which complement those obtained from other
NC observables.

The prospects for achieving such improved precision are promising. In
the case of cesium atomic PV, the systematic error is already at the
0.5\% level, and one anticipates achieving a total experimental error
of 0.5\% in the future. The present atomic theory error is 1.2\%, and
it remains to be seen whether atomic theorists can push this error
below one percent in the future. Previous experience with the MIT-Bates
$^{12}$C PV electron scattering experiment indicates that achieving
systematic error on the order of 1\% is within reason, while the high
luminosity available at TJNAF implies that obtaining a similar level
of statistical error is possible with an experiment of realistic running
time. 
Moreover, the present PV program at MIT-Bates, TJNAF, and MAMI should
yield sufficient information on the NC form factors $F(Q^2)$ appearing
in Eq. (6) to render them a negligible source of uncertainty for
a $Q_W$ extraction. The primary theoretical challenge appears to
be obtaining a realistic evaluation of the uncertainty in $\Delta
Q_W(\hbox{had})$ for this process.

\medskip
\noindent{\bf Nuclear Parity Violation}
\medskip

From the standpoint of electroweak theory, nuclear PV observables
are of interest as a window on the $\Delta S=0$ hadronic weak
interaction. Within the framework of the SM, this interaction 
is composed of $I=0,1,2$ components. One may correspondingly
write down a two-body nuclear Hamiltonian having the same isospin
content:
\begin{equation}
{\hat H}^{PV}_{NN} = \sum_i h_i {\hat{\cal O}}_i(2) \ \ \ ,
\end{equation}
where the $h_i$ are constants dependent on the hadronic weak
interaction, the ${\hat{\cal O}}_i(2)$ are two-body nuclear operators
containing various spin, isospin, and momentum structures, and the
index $i$ runs over the possible channels containing $I=0,1,2$. Given
the hard core of the strong $NN$ potential, it is un-likely that
the weak NN force is mediated by the exchange of a weak vector boson
between two nucleons. Rather, one expects the exchange of mesons
to dominate the weak two-nucleon potential. Under this {\em ansatz},
the $h_i$ are given as products of strong and weak PV meson-nucleon
couplings:
\begin{equation}
h_i=g^i_{NNM}\, h^i_{NNM} \ \ \ ,
\end{equation}
where the $h^i_{NNM}$ ($g^i_{NNM}$) are weak PV (strong) meson ($M$)
nucleon ($N$) couplings. In order to obtain all of the isospin components
required by the structure of the $\Delta S=0$ quark-quark interaction,
one must include the exchange of the $\pi$, $\rho$, and $\omega$ mesons
at a minimum. 

When seeking to extract information on the hadronic weak interaction from
PV nuclear observables, one must undertake several levels of analysis. From
a phenomenological standpoint, the problem is to determine whether one
may obtain a consistent set of $h_i$ from a global analysis of observables.
In this respect, one must also rely upon nuclear theory to provide
computations of nuclear matrix elements 
(such as $\langle A| {\hat{\cal O}}_i(2)
|A\rangle$) in order to extract the $h_i$ from experimental quantities.
Within the framework of the meson-exchange picture, one would also
like to understand how the values of the $h^i_{NNM}$ arise from the PV
four-quark weak Hamiltonian, ${\hat{\cal H}}_{PV}$. This objective
presents hadron structure theorists with the problem of reliably computing
weak matrix elements $\langle NM| {\hat{\cal H}}_{PV}| N\rangle$, a non-trivial
task. In fact, there has been little progress in this direction since
the \lq\lq benchmark" quark model calculation of Ref. [9] more than 15
years ago. Clearly, deepening our understanding of the $\Delta S=0$
hadronic sector of the SM requires progress on a variety of fronts.

To date, most experimental information has been derived from two broad
classes of observables: so-called \lq\lq direct" $NN$ studies, such as
${\vec p}+ p$ scattering, and studies of PV in light nuclei, such as
the PV $\gamma$ decays of $^{18}$F and $^{19}$F. In the latter instance,
the size of the PV observable is enhanced by the mixing of nearly-degenerate
opposite parity states by ${\hat H}^{PV}_{NN}$. In both direct and
light-nuclei studies, the observables involve some form of spin polarization.
At present, one has yet to achieve a consistent set of the $h_i$
from a rather broad sample of observables. Of particular interest is
the constraint from $^{18}$F PV on $h_{NN\pi}$, which is in conflict
with constraints from ${\vec p}+\alpha$, $^{19}$F, and $^{21}$Ne
experiments [10]. The reason for this discrepancy is not fully understood,
but the possibilities include (a) an error in one of the experiments
(b) significant nuclear theory uncertainty in the computation of PV
nuclear matrix elements, or (c) the omission of important terms
from the model of Eq. (7). A recent analysis of Brown and collaborators
[11], performed in a truncated model space but including higher
$n\hbar\omega$ configurations shift the $^{21}$Ne constraints in such
a manner as to bring all the bounds on $h_{NN\pi}$ into agreement.
Whether this trend emerges from a computation using a complete
model space remains to be seen [12].

The measurement of new observables is clearly desirable, as such 
measurements could yield new and, in principle, complementary constraints
on the $h_i$. In this respect, two possibilities have received considerable
interest recently: scattering of epithermal polarized neutrons from heavy
nuclei [13] and observation of the nuclear anapole moment via atomic PV 
experiments [14]. In the case of the neutron scattering experiments, one
measures a neutron transmission asymmetry associated with incident neutrons
of opposite helicity. Such an asymmetry arises from the mixing of $s$-wave
resonances into $p$-wave resonances by the interaction of Eq. (7). Due to
the large level densities for nuclei such as $^{238}$U, the energy denominators
associated with the mixing are small. Moreover, the $s$-wave resonances
couple strongly to the continuum. As a consequence of these two features,
the PV asymmetry can be signficantly enhanced. From a theoretical standpoint,
the presence of a large number of neighboring $s$- and $p$-wave resonances
complicates the analysis, and one must resort to statistical approaches
in extracting information about the PV nuclear force. To the extent that
the assumptions of the statistical models are valid, one derives from
the asymmetry a mean square PV nuclear matrix element. The corresponding
constraints on the $h_i$ take the shape of a quadratic form in the 
multi-dimensional space of PV couplings. These constraints appear to
be consistent with the constraints on $h_{NN\pi}$ derived from the
$^{18}$F and $^{21}$Ne $\gamma$-decays but not with
the constraints obtained from $^{19}$F or ${\vec p}+\alpha$ experiments.
In the case of one nucleus, $^{232}$Th, the results display a deviation
from the pattern expected within the conventional statistical model.
The mean value of the measured $^{232}$Th asymmetries differs from
zero by more than two standard deviations (one expects this average to
be zero in the statistical model). Various extensions of the statistical
approach used to account for this deviation yield a mean value for the
PV matrix element which is two orders of magnitude larger than the scale
of PV matrix elements implied by other PV measurements. While this
so-called \lq\lq sign problem" arises only in the case of $^{232}$Th, 
one has yet to obtain a satisfactory explanation.

The second new approach to placing new constraints on the $h_i$ is
to measure the nuclear anapole moment (AM). The AM is an axial
vector coupling of the photon to the nucleus induced by parity-mixing
in the nucleus. Technically, it is an elastic matrix element of the
transverse electric multipole operator, ${\hat T}^E_{J=1}$ -- a matrix
element which must vanish in the absence of PV. This matrix element 
goes like $Q^2$ for small momentum transfer. Consequently, the
AM couples only to virtual photons, such as those exchanged between
the nucleus and atomic electrons. Moreover, because the leading $Q^2$ of the
AM cancels the $1/Q^2$ of the photon propagator, the corresponding
interaction is contact-like in co-ordinate space. In this respect, the 
contribution made by the AM to PV observables is indistinguishable from
the $V(e)\times A(N)$ NC interaction. Indeed, both induce a 
nuclear spin-dependent (NSD) atomic interaction of the form
\begin{equation}
{\cal H}^{PV}_{ATOM}(\hbox{NSD}) = {G_F\over 2\sqrt 2}\ {\tilde k}
\ \psi^{\dag}_e(0) {\vec \alpha} \psi_e(0)\cdot {\vec I} \ \ \ ,
\end{equation}
where $\psi_e$ is the electron field, $\vec\alpha$ is the vector of
Dirac matrices, $\vec I$ is the nuclear spin, and the constant
$\tilde k$ can be decomposed into a sum of NC and AM contributions:
${\tilde k}_{NC}+{\tilde k}_{AM}$. 

Although ${\tilde k}_{AM}$ is nominally suppressed with respect to
${\tilde k}_{NC}$ by a factor of $\alpha$, one may nevertheless expect
it to make an observable contribution for the following two reasons.
First, the NC contribution is suppressed since ${\tilde k}_{NC}\propto
g_V^e=-1+4\sin^2\theta_W\approx -0.1$. Second, the scale of the AM
grows with the square of the nuclear radius, and thus as $A^{2/3}$,
whereas NC contribution receives no coherence enhancement. Hence,
for heavy nuclei, ${\tilde k}_{AM}/{\tilde k}_{NC}$ can be as large
as three or more, according to a variety of calculations [15]. Although 
recent results from the cesium atomic PV experiment
is not conclusive, it is nevertheless consistent with theoretical
expectations [16]. The result from the atomic thallium experiment [16]
differs from theoretical predictions by about $2\sigma$ and has the
opposite sign. There is undoubtedly room for improvement on the
theoretical side as well as on the part of experiment. Ideally,
future measurements of NSD atomic PV observables will achieve 
significantly better precision and, coupled with theoretical progress,
yield new constraints on the $h_i$.

\medskip
\noindent{\bf Parity Conserving Time-reversal Violation}
\medskip

Finally, I wish to touch briefly on a subject which has received
renewed interest recently -- searches for parity conserving
time-reversal violating (PCTV) interactions. Traditionally, such 
searches have relied on three classes of studies: detailed
balance in nuclear reactions, nuclear $\gamma$-decays, and 
$\beta$-decay [17]. More recently, constraints on PCTV physics have
been derived from two other observables: the five-fold correlation
in epithermal neutron scattering from heavy nuclei
and the permanent electric dipole
moments (EDM's) of atoms and nuclei. In the case of neutron scattering,
one looks for a scattering phase shift $\delta_{PCTV}$ proportional to 
\begin{equation}
{\vec s}\cdot({\vec I}\times{\vec p})({\vec I}\cdot{\vec p})
\end{equation}
where $\vec s$ is the spin of an incident neutron of momentum $\vec p$
and $\vec I$ is the nuclear spin. Typically, searches for EDM's try
to detect a frequency shift $\Delta\nu\sim d{\vec J}\cdot{\vec E}$,
where $d$ is the EDM, $\vec E$ is a static, applied electric field,
and $\vec J$ is the spin of the quantum system of interest. 

What makes $\delta_{PCTV}$ and $d$ particularly interesting is their
sensitivity to PCTV flavor-conserving ($\Delta F=0$)
interactions. Within the context
of renormalizable gauge theories, $\Delta F=0$ PCTV interactions
between quarks cannot arise at ${\cal O}(g^2)$, where $g$ 
is the gauge coupling [18].
In order to generate them, one requires loops involving gauge interactions
which exist beyond the framework of the SM. Alternatively, one may
work with non-renormalizable effective interactions which apply below
some scale $M_X$ associated with PCTV gauge interactions. In the latter
instance, one finds a class of dimension seven operators which can
generate a $\delta_{PCTV}$ or EDM at the quark level [19]. The list of such
operators includes, for example,
\begin{eqnarray}
{\beta\over M_X^3} \bar{\psi}\gamma_5 {\buildrel\leftrightarrow\over D}_\mu 
	\psi \bar{\psi}\gamma_5\gamma_\mu \psi \\
{eg{\tilde\beta}\over M_X^3}\bar{\psi}\sigma_{\mu\lambda}\psi 
	F^{\rho\mu}_{(\gamma)} F^{\lambda\nu}_{(Z)} g_{\rho\nu}
\end{eqnarray}
where $\psi$ is a fermion field and $F^{\mu\nu}_{(\gamma)}$
($F^{\mu\nu}_{(Z)}$) is the field strength associated with the
photon ($Z$-boson). Both of these interactions may generate contributions
to the EDM of a quark. To do so, they require the presence of a PV
weak interaction, since the interaction of an EDM with a static electric
field is both parity and time-reversal violating. Assuming, for illustrative
purposes, that the couplings $\beta$ and $\tilde\beta$ are of order unity,
the present limits on the neutron EDM would constrain the mass scale
$M_X$ to be roughly two orders of magnitude larger than $M_Z$ [20].

An alternate scheme for treating PCTV in the purely hadronic sector is
to employ a meson-exchange model. In this case, the lightest allowed
meson is the $\rho$, which can interact with one nucleon through the
PC strong interaction and the other nucleon through a PCTV interaction. 
The associated PCTV coupling is conventionally denoted $\bar{g}_\rho$.
The present upper bounds on $\bar{g}_\rho$ from detailed balance and
$\delta_{PCTV}$ are about 2.5 and 22, respectively [18]. One expects the
epithermal neutron bounds to improve by a factor of 100 or so with the
completion of future measurements. Alternately, one may derive limits
on $\bar{g}_\rho$ from atomic EDM's by assuming that the EDM is generated
by (a) PCTV in the purely hadronic sector and (b) an additional PV
weak interaction either inside the nucleus or between the nucleus and
atomic electrons. In the case of the neutron EDM, all of the symmetry
violating interactions are hadronic. The corresponding limits on
$\bar{g}_\rho$  obtained from the neutron EDM are roughly $< 10^{-3}$,
while those obtained from atomic EDM's are about an order of magnitude
weaker [21]. With the advent of more precise atomic EDM measurements, one
would anticipate deriving better limits on $\bar{g}_\rho$. The theoretical
problem of understanding how such bounds would translate into constraints
on PCTV quark-quark interactions remains open.

\medskip
\noindent{\bf Conclusions}
\medskip

In this talk, I hope to have convinced you that the use of spin-dependent
atomic and nuclear weak interaction observables have an important and
on-going role to play in searching for electroweak physics  beyond the
Standard Model. In the case of semi-leptonic PV, for example, the prospects
are good for placing bounds on the scale of compositeness and the mass of
an additional neutral gauge boson which are competitive with, or better
than, those one might achieve with high-energy acclerators. Similarly,
the study of hadronic parity violation continues to challenge 
our understanding of the purely hadronic weak interaction. Finally,
the analysis of spin observables such as the five-fold correlation in
epithermal neutron scattering or neutron and atomic EDM's yield constraints
on the scale of PCTV interactions which may arise in certain extensions of the
SM. Undoubtedly, the continuing improvement in the precision with which
spin-dependent atomic and nuclear observables are measured will provide
an abundant supply of grist for the electroweak theorist's mill. 

\medskip
\noindent{\bf Acknowledgements}
\medskip
 
This work was supported in part under U.S. Department of Energy
contract \# DE-FG06-90ER40561 and a National Science Foundation
Young Investigator Award.

\vspace{0.2cm}
\vfill
{\small\begin{description}
\item{[1]}
D. London and J.L. Rosner, Phys. Rev. {\bf D34} (1986) 1530.
\item{[2]}
M.K. Pillai {\em et al.}, CDF Collaboration, meeting of the Division
of Particles and Fields, American Physical Society, August, 1996.
\item{[3]}
E. Eichten, K. Lane, M.E. Peskin, Phys. Rev. Lett. {\b 50} (1983) 811;
P. Langacker, M. Luo, A. Mann, Rev. Mod. Phys. {\bf 64} (1992) 87.
\item{[4]}
A. Bodek, CDF Collaboration, ICHEP Conference, Warsaw, July 1996.
\item{[5]}
M.E. Peskin and T. Takeuchi, Phys. Rev. Lett. {\bf 65} (1990) 964;
W.J. Marciano and J.L. Rosner, Phys. Rev. Lett. {\bf 65} (1990) 2963.
\item{[6]}
J.L. Rosner, Phys. Rev. {\bf D53} (1996) 2724.
\item{[7]}
B.P. Masterson and C.E. Wieman in {\sl Precision Tests of the Standard
Electroweak Model}, P. Langacker ed., World Scientific, Singapore, 1995.
\item{[8]}
M.J. Musolf {\em et al.}, Phys. Rep. {\bf 239} (1994) 1.
\item{[9]}
B. Desplanques, J.F. Donoghue, B.R. Holstein, Ann. Phys. (NY) {\bf 124}
(1980) 449.
\item{[10]}
See, {\em e.g.}, B.R. Holstein in {\sl Symmetries and Fundamental Interactions
in Nuclei}, W.C. Haxton and E.M. Henley eds., World Scientific, Singapore,
1995.
\item{[11]}
M. Hiroi and B.A. Brown, Phys. Rev. Lett. {\bf 74} (1995) 231.
\item{[12]}
W.C. Haxton, private communication.
\item{[13]}
J.D. Bowman, G.T. Garvey, and M.B. Johnson, Ann. Rev. Nucl. Part. Sci. 
{\bf 43} (1993) 829.
\item{[14]}
M.J. Musolf and B.R. Holstein, Phys. Rev. {\bf D43} (1991) 2956.
\item{[15]}
V.V. Flambaum, I.B. Khriplovich, O.P. Sushkov, Phys. Lett. {\bf B146}
(1984) 367; W.C. Haxton, E.M. Henley, M.J. Musolf, Phys. Rev. Lett.
{\bf 63} (1989) 949; C. Bouchiat and C.A. Piketty, Z. Phys. C {\bf 49}
(1991) 91.
\item{[16]}
M.C. Noeker, {\em et al.}, Phys. Rev. Lett. {\bf 61} (1988) 310; P.A.
Vetter, {\em et al.}, Phys. Rev. Lett. {\bf 74} (1995) 2658.
\item{[17]}
See, {\em e.g.}, F. Boehm in {\sl Symmetries and Fundamental Interactions
in Nuclei}, {\em Op. Cit}.
\item{[18]}
P. Herczeg in {\sl Symmetries and Fundamental Interactions in Nuclei}, 
{\em Op. Cit.}.
\item{[19]}
I.B. Khriplovich, Nucl. Phys. {\bf B352} (1991) 385; J. Engle, P.H.
Frampton, R. Springer, Phys. Rev. {\bf D53} (1996) 1542.
\item{[20]}
M.J. Musolf and I. Grigentch, in preparation.
\item{[21]}
W.C. Haxton, A. H\"oring, M.J. Musolf, Phys. Rev. {\bf D50} (1994) 3422.
\end{description}

\vfill
\eject
\end{document}